\documentstyle[osa,manuscript,epsbox]{revtex}  

\begin{document}                

\title{History-dependent Maximum Static Friction and Rectification of a System with a Few Particles in a Periodic Field}

\author{Akinori Awazu \footnote{E-mail: awa@complex.c.u-tokyo.ac.jp} }

\address{Department of Pure and Applied Sciences University of Tokyo,\\
Komaba 3-8-1, Meguro-ku, Tokyo 153-8902, Japan.} %

\maketitle
\begin{abstract}
History-dependent maximum static friction is investigated using simple 
models with two and three particles in one-dimensional periodic 
potentials. In some situations, these systems possess two values of the 
maximum static friction,
with that actually realized being determined by the direction
of the slippage relative to the direction of the previous slippage. 
In particular, the maximum static friction for slippage in a given 
direction is smaller when the previous slippage was in the same direction 
than when it was in the opposite direction.
Owing to this property, a particle can continue to move in a direction even 
in the case that it is subject to an external force whose direction change 
in time as in a ratchet 
system which is regarded as one of the simplest model of molecular motors.
\vspace{2mm}
PACS number(s):\\

\end{abstract}

Static and dynamic frictions and the
transition between them are universal phenomena commonly observed at 
the surfaces of macroscopic 
objects\cite{sato00,sato00a,sato00b,robin,sato0,sato001,nasuno,nasunoa,
hayakawab,hayakawa,sato1,sato1a,awa}. 
There is a rich variety of such phenomena, and their characteristics
depend on the physical properties of the surfaces as well as any 
lubricants that might exist between them\cite{nasuno,nasunoa,hayakawab,hayakawa,sato1,
sato1a}. 
In some cases, frictions can depend on the system history
\cite{sato0,sato001,nasuno,nasunoa}. For example, the maximum static 
friction can depend on how long the two surfaces have been in contact
\cite{sato0}. 
Also, in some case, dynamic friction with increasing slippage velocity 
is larger than 
that with decreasing velocity\cite{sato001,nasuno,nasunoa}. 

In this paper, the existence of history-dependent maximum static 
friction is investigated using simple systems. First, a system 
consisting of only two particles in a spatially periodic field is considered. 
In this system, the maximum static friction depends on the direction of the 
previous slippage. We show that this property can cause the system to 
exhibit rectification behavior.
We also show that similar phenomena are observed in a system consisting 
of three particles in a spatially periodic field.

We study a one dimensional system containing two particles in which both 
particles are subject to a spatially periodic potential and one particle 
is subject to an external force $F_{ex}$ (Fig. 1).
The motion of these two particles is described by  the over-damped equations,
\begin{equation}
\dot{x_{1}} = c_{1}c_{2}\sin(2 \pi (x_{1}-x_{2}))+c_{1}c_{p}\sin(2 \pi x_{1})+F_{ex}
\end{equation}	
\begin{equation}
\dot{x_{2}} = c_{1}c_{2}\sin(2 \pi (x_{2}-x_{1}))+ c_{2}c_{p}\sin(2 \pi x_{2})
\end{equation}
where $x_{i}$ is the $i$th particle's position. 
We set $c_{1}=c_{p}=1$ and 
$c_{2}=c > 0$, and we consider the system to be defined for $0 \le x_i <1$, 
with periodic boundary conditions. 
This system belongs to a class of coupled phase oscillators systems that 
have been studied extensively\cite{kaneko,kuramoto,daido}.

The above system can be regarded as a simplified model of a physical system  
in which thin lubricants is spread uniformly between 
two objects with bumpy surfaces, and an external force acts on one of the 
two objects. 
In this situation, the motion of the contact points of the two objects and 
the motion of the lubricant particles 
are, approximately, modeled by the motion of the first and second 
particle, as described by Eqs. (1) and (2), respectively (Fig. 1). 
The condition $c_{i}>0$ means that, as the interactions in the system, we 
consider only the effects of repulsive forces like the excluded 
volume effect which plays important roles in liquids, solids and gels.

In the following, we report the results of simulations employing above 
system whose purpose is to determine the relationships 
between the maximum static friction and observed microscopic states. 
We use $F_{ex} = (1-\cos (t/T))/2$ with large $T$, so that 
the external force varies smoothly and very slowly. The data points 
($\times$) in Fig. 2 indicate values of the maximum static friction $R$, 
plotted as a function of $c$ $(0 \le c \le 1)$ for two particle 
configurations, $x_{1}^{b} > x_{2}^{b}$ and $x_{1}^{b} < x_{2}^{b}$, before 
slippage. Here, $R$ is defined as the maximum value of $|F_{ex}|$ 
for which the first particle remains stuck as $|F_{ex}|$ increases, and 
$x_{i}^{b}$ is the $i$th particle's position at the most recent time at which 
$F_{ex}=0$. This $c$ - $R$ relation is divided into three regions: I) for 
$c < \hat{c}_{crit^{1}} \approx 0.59$, $R$ is decreasing function of $c$ 
for all $x_{1}^{b}$, $x_{2}^{b}$, II) for 
$\hat{c}_{crit^{1}} < c < \hat{c}_{crit^{2}} \approx 0.66$, $R$ has two 
possible 
values, one realized for $x_{1}^{b} > x_{2}^{b}$ and one for 
$x_{1}^{b} < x_{2}^{b}$, and III) for $c > \hat{c}_{crit^{2}} $, $R$ is 
an increasing function of $c$ for all $x_{1}^{b}$, $x_{2}^{b}$. 

When $|F_{ex}(t)|$ decreases from some initial value greater than $R$, 
the first particle ceases slipping at a value of $|F_{ex}|=R_{stop}$.
Through our simulations, we found relations between $R_{stop}$ and $R$ for 
each of the above described cases: $R_{stop}=R$ 
in case I), $R_{stop} < R$ in case III) as shown in Fig. 2, and 
$R_{stop}=R_{smaller}$ in case II), where $R_{smaller}$ is the smaller 
of the two values of $R$.

Figure 3 displays typical temporal evolutions of 
each particle's velocity and position for cases I) and III), with 
$F_{ex}$ as given above. Here, in (a) $c=0.2$ and $x_{1}^{b} < x_{2}^{b}$, 
in (b) $c=0.2$ and $x_{1}^{b} > x_{2}^{b}$, in (c) $c=0.8$ and 
$x_{1}^{b} < x_{2}^{b}$, and in (d) $c=0.8$ and 
$x_{1}^{b} > x_{2}^{b}$. In this figure, the gray curves represent position
and velocity of the first particle, and the black curves represent those of 
the second. 
In the situation depicted in Fig. 3 (a) and (b), $x_{1} > x_{2}$ always 
holds just before the slippage of the first particle, independently of 
$x_{1}^{b}$ and $x_{2}^{b}$. 
This is due to the fact that if $F_{ex}$ is small, for sufficiently small 
$c_{2}$, it can be the case that the first particle is not able to cross the 
potential barrier, while it is able to cross the second particle. 
In fact this is the case in the situations described by Figs. 3(a) and (b).
Contrastingly, in Figs. 3 (c) and (d), $x_{1} < x_{2}$ always holds just 
before the slippage of the first particle, independently of $x_{1}^{b}$ and $x_{2}^{b}$. 
This follows from the fact that if $F_{ex}$ is small, for sufficiently 
large $c_{2}$,
it can be the case that the first particle cannot cross the second particle, 
while it can cross the potential barrier.
From these considerations, it is clear why $R$ has only a single value for 
each $c$ value in cases of I) and III).

By considering the balance equations obtained by setting 
$\dot{x_{1}}=0$ and $\dot{x_{2}}=0$ in Eqs. (1) and (2) and choosing the 
particle configuration before the slippage, we can obtain $c$-$R$ curves 
plotted in Fig. 2. 
For example, the curve of $R$ obtained as the maximum values 
of $F_{ex}$ with $\dot{x_{1}}=0$ and $\dot{x_{2}}=0$ under the condition 
$x_{1}>x_{2}$ is consistent with the numerical result for 
$c < \hat{c}_{crit^{2}}$, where $R = -[\sin(2X)+c \sin(X)]$ with 
$X=\arccos[(-c-(c^{2}+32)^{1/2})/8]$.
The line $R = c$ is consistent with the numerical results for 
$c > \hat{c}_{crit^{1}}$, which is obtained from 
$R = -[c \sin(2 \pi (x_{1}-x_{2})) + \sin(2 \pi x_{1})]|_{x_{1} = 0.5, x_{2} = 0.75}$. 
Here $x_{1} = 0.5$ and $x_{2} = 0.75$ corresponds to the values for which 
the force of the second particle on the first under the condition 
$x_{1}<x_{2}$ is maximal. 

In case II), the maximum static friction is determined by the 
history, which direction did the first particle slip previously.
This can be understood as follows.
Figure 4 displays typical temporal evolutions of each particle's velocity and 
position for case II). (Here, $c=0.63$ and in (a) $x_{1}^{b} < x_{2}^{b}$ 
and in (b) $x_{1}^{b} > x_{2}^{b}$.) In contrast to cases I) and III), here, 
the particles do not cross as $F_{ex}$ is increased until the first particle 
starts to slip. We also found that slippage begins at a later time for 
$x_{1}^{b} < x_{2}^{b}$ than for $x_{1}^{b} > x_{2}^{b}$ . 
This means that $R$ depends on 
the particle configuration: $R$ for $x_{1}^{b} < x_{2}^{b}$ is larger than 
that for $x_{1}^{b} > x_{2}^{b}$. Moreover, the state with 
$x_{1} > x_{2}$ is realized after slippage of the first 
particle whether $x_1^b > x_2^b$ or $x_1^b < x_2^b$. 
(Because of the symmetry of this system, $x_{1} < x_{2}$ is 
necessary realized after slippage of the first particle if $F_{ex}<0$.)
This means that $R$ for slippage in the same direction as the previous 
slippage is always smaller than that for the slippage in the opposite 
direction. 

If $F_{ex}>0 $ and it varies slowly and smoothly in time, in case II), 
$x_{1} > x_{2}$ is realized after slippage. We now explain the reason for 
this. Figure  4 (c) plots the temporal evolutions of each 
particle's velocity and position for $c=0.63$ with $F_{ex} = 0.583$, which 
is slightly larger than $R_{smaller}$. 
If $F_{ex}$ is constant and 
$|F_{ex}| > R $, the first particle slips and the second particle oscillates 
periodically. As shown in Fig. 4 (c), the time required to switch 
from $x_1 > x_2$ to $x_1 < x_2$ is much longer than that to switch from 
$x_1 < x_2$ to $x_1 > x_2$ when $F_{ex}$ is slightly larger than 
$R_{smaller}$. If $c$ is not large, the amplitude of the 
second particle's oscillation is not large. 
In such a situation, $x_{2}$ cannot reach such a value ($x_{2} \sim 0.75$) 
that the force of the second particle on the first is large enough to 
balance $F_{ex}$ at $x_{1} < x_{2}$. 
Hence, the system cannot remain in a state with $x_1 < x_2$ for an 
extended time if the first particle crosses over the potential barrier. 
in contrast to case III) depicted in Figs. 3 (c) and (d). 

Also in case III), the second particle cannot reach at $x_{2} \sim 0.75$ 
when $|F_{ex}|$ decreases from a value greater than $R$. 
However, in this case, the $x_{2}$ can reach such a value that the force of 
the second particle on the first at $x_{1} < x_{2}$ balances $F_{ex}$ with 
$|F_{ex}| = R_{stop}$, in contrast to case II).

Since, in case II), the static maximum friction to the direction as the 
direction of previous slippage is smaller than that to the opposite 
direction, for this case, this system exhibits rectification behavior.
Recently, rectification behavior has been observed, for example, in some 
ratchet systems that are simple models of molecular 
motors\cite{moter,sekimoto,prost}. 
In these models, with an external force that favors neither direction, 
particles can move in only 
one direction, which is completely determined by the shape of the potential. 
Now, to demonstrate that our model can exhibit similar behavior, 
we consider the case $c=0.63$ and $F_{ex} = 0.61 \sin(t/T)$, with 
sufficient large $T$. Thus, in this case, the 
external force acts symmetrically in both positive and negative directions. 
Here, we choose the value $0.61$ because it is halfway between the two 
values of $R$ for $c=0.63$. For this system, as shown in the left-hand side 
of Fig. 4 (d), the first 
particle can move only in one direction, even though $F_{ex}$ alternates 
between positive and negative values. 
However, if a sufficient strong force is applied instantaneously to the 
first particle in the direction opposite to its motion, its motion can 
reverse direction [see the right-hand side of Fig. 4 (d)]. 
This means we can control the slippage direction in the system. This is not 
possible in the ratchet systems proposed as models of molecular motors.

The phenomena discussed above can also be observed in systems with a 
larger number of degrees of freedom. As an example, we consider a system 
that consists of three particles in a spatially periodic field, with the 
motions of each particle obeying the equation,
\begin{equation}
\dot{x_{i}} = \sum_{i \ne j}c_{i}c_{j}\frac{1+\cos(2 \pi (x_{i}-x_{j}))}{2}\sin(2 \pi (x_{i}-x_{j}))+c_{i}c_{p}\frac{1+\cos(2 \pi x_{i})}{2}\sin(2 \pi x_{i})+\delta_{i,1}F_{ex}.
\end{equation}
The characteristic length of the interactions between particles in this 
system is shorter than that of the system described by (0.1) and (0.2). 
Again, we consider the system to be defined in the region $0 \le x_{i} <1$ 
and use periodic boundary conditions.
In the case that $c_{1}=c_{2}=1$ and $c_{3}=c_{p}=c$, behavior similar to 
that exhibited in cases II) for the two-particle system is observed over 
a wide range of values of $c$
with $F_{ex} = F (1-\cos (t/T))/2$, for $F>0$ and sufficiently large $T$. 

The solid curves in Fig. 5 represent $R$ as a function of $c$ 
$(0.1 \le c \le 1)$. As shown, $R$ takes two values 
for each value of $c$, depending on the relationships among the $x_{i}^{b}$. 
Figures 6 (a), (b) and (c) displays typical temporal evolutions of each 
particle's velocity and position with $c=0.5$ for two particle 
configurations before slippage; the (1,2,3) configuration in which the 
particles are arranged in the order $1$, $2$, $3$ with respect to the 
direction of $F_{ex}$ (the direction of increasing $x$ in this case), as 
in (a), and the (1,3,2) configuration, as in (b) and (c). 
Here, $F=0.5$ in (a) and (b), 
and $F=0.36$ in (c). By comparing (a) and (b), we find that $R$ for 
the (1,3,2) configuration is larger than that for the (1,2,3) configuration. 
Also it is seen that the (1,2,3) configuration is always realized 
after slippage in both cases considered in (a) and (b). 
We are thus led to the conclusion that, as in the system with two particles, 
under a certain condition, $R$ for the direction of the previous slippage is 
always smaller than $R$ for the direction opposite to the previous slippage.
For the three particle system, this condition is that $F$ be larger than 
a particular value, which we discuss below.

As stated above, the (1,2,3) configuration
is apparently always realized after slippage in the situations considered 
in Figs. 5 (a) and (b).
If $F$ is not too large, however, the (1,3,2) configuration
can be preserved upon slippage. In fact, this is the case  for the situation 
considered in Fig. 6 (c).
The dotted curve in Fig. 5 represent critical 
values of $F_{ex}$, below which the particle configuration is preserved 
upon slippage. 
Thus, in this system, the maximum strength of the external force at the 
slippage determines whether or not there is a direction dependence of $R$.
We found that with a properly chosen $F_{ex}$, this system too can exhibit 
rectification behavior. 

In this paper, we have investigated the history dependence of the maximum 
static friction using simple systems consisting of two and three 
particles in a one-dimensional periodic potential. In these system, 
we found that, in some cases, the maximum static friction can depend on 
the direction of the slippage, being smaller than this direction is the 
same as that of the previous slippage. 
By this property, a particle in this system can continue to slip along a 
single direction even in the case that the direction of the external force 
change in time. This behavior is similar to that seen in ratchet systems 
regarded as one of simplest models of molecular motors. 
Analytical study of the systems considered here 
and further investigation of systems with three or more
particles will be carried out in the future. 
The construction of a model that can realize more types 
of memory effects, for example aging of friction\cite{sato0}, is also an 
important future problem.

The author is grateful to K. Kaneko, K. Fujimoto, K. Sekimoto and H. Matsukawa 
for useful discussions. This research was supported in part by a Grant-in-Aid 
for JSPS Fellows (10376).

\newpage

\begin{figure}[ht]
\begin{center}
\end{center}
\caption{Schematic depiction of the system described by (0.1) and (0.2), consisting of two particles in a spatially periodic potential.}
\end{figure}

\begin{figure}[ht]
\begin{center}
\end{center}
\caption{($\times$) points and dotted lines represent maximum value of the 
static friction of the system $R$ as a 
function of $c$ for two conditions, $x_1^b > x_2^b$ and $x_1^b < x_2^b$, 
before slippage. The ($\times$) points are the results of our simulation, and 
dotted lines were obtained analytically. (+) points represent $R_{stop}$ as a 
function of $c$ in case III).}
\end{figure}

\begin{figure}[ht]
\begin{center}
\end{center}
\caption{Typical temporal evolutions of the velocity and position of each 
particle for case with I) in (a) and (b) and case III) in (c) and (d) with 
slowly changing of $F_{ex}$. In (a) $c=0.2$ and $x_{1}^{b} < x_{2}^{b}$, 
in (b) $c=0.2$ and $x_{1}^{b} > x_{2}^{b}$, in (c) $c=0.8$ 
and $x_{1}^{b} < x_{2}^{b}$, and in (d) $c=0.8$ and 
$x_{1}^{b} > x_{2}^{b}$. The gray curves represent the first particle, 
and the black curves represent the second particle. The thickness of 
each curve is proportional to the value of $c_i$ for the particle to 
which it corresponds. The thin sinusoidal curve represents
$F_{ex}$. The numbers at the right of the figures are the particle label.}
\end{figure}

\begin{figure}[ht]
\begin{center}
\end{center}
\caption{Typical temporal evolutions of the velocity and position of each 
particle for case II) with a slowly changing $F_{ex}$ in (a) and (b), a static $F_{ex}$ with the value $0.583$ which is just larger than $R_{smaller}$ 
in (c), and a slowly changing $F_{ex} = 0.61 \sin(t/T)$ in (d). 
In (a) $c=0.63$ and $x_{1}^{b} < x_{2}^{b}$, and in (b),(c) and (d) 
$c=0.63$ and $x_{1}^{b} > x_{2}^{b}$. Shades and widths of the curves, and 
the numbers at the right of the figures have the same meanings as in Fig. 3.}
\end{figure}

\begin{figure}[ht]
\begin{center}
\end{center}
\caption{The solid curves represent typical maximum values of the static 
friction $R$ as the function of $c$ in the system consisting of three 
particles for two configurations, the (1,2,3) configuration (lower curve) 
and the (1,3,2) configuration (upper curve) before slippage. 
The dotted curve represents the critical values of $F_{ex}$, for which 
the (1,3,2) is preserved under slippage.}
\end{figure}

\begin{figure}[ht]
\begin{center}
\end{center}
\caption{Typical temporal evolutions of the velocity and position of each 
particle in the three particle system with $c=0.5$.
The shades and widths of the curves have the same meanings as in Fig. 3.
The external forces $F_{ex}$ used here are $F_{ex} = (1-\cos (t/T))/2$ in 
(a), (b), and (c).
The particle configuration before slippage is (1,2,3) in (a) 
and (1,3,2) in (b) and (c).}
\end{figure}

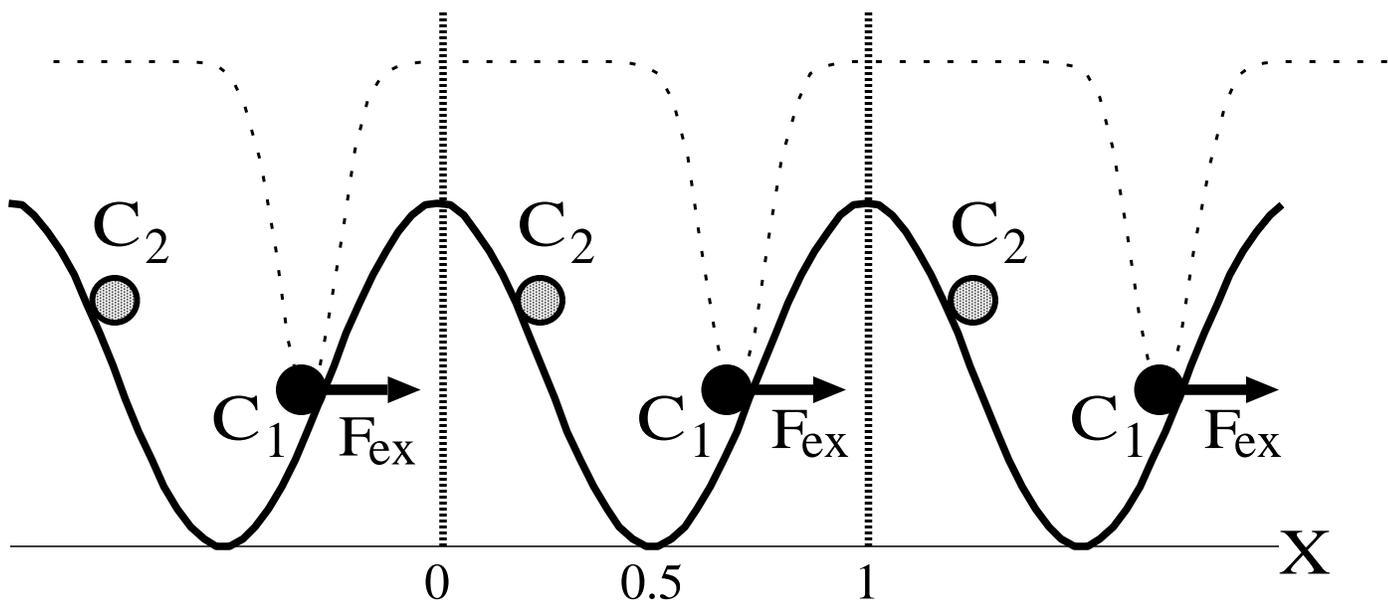
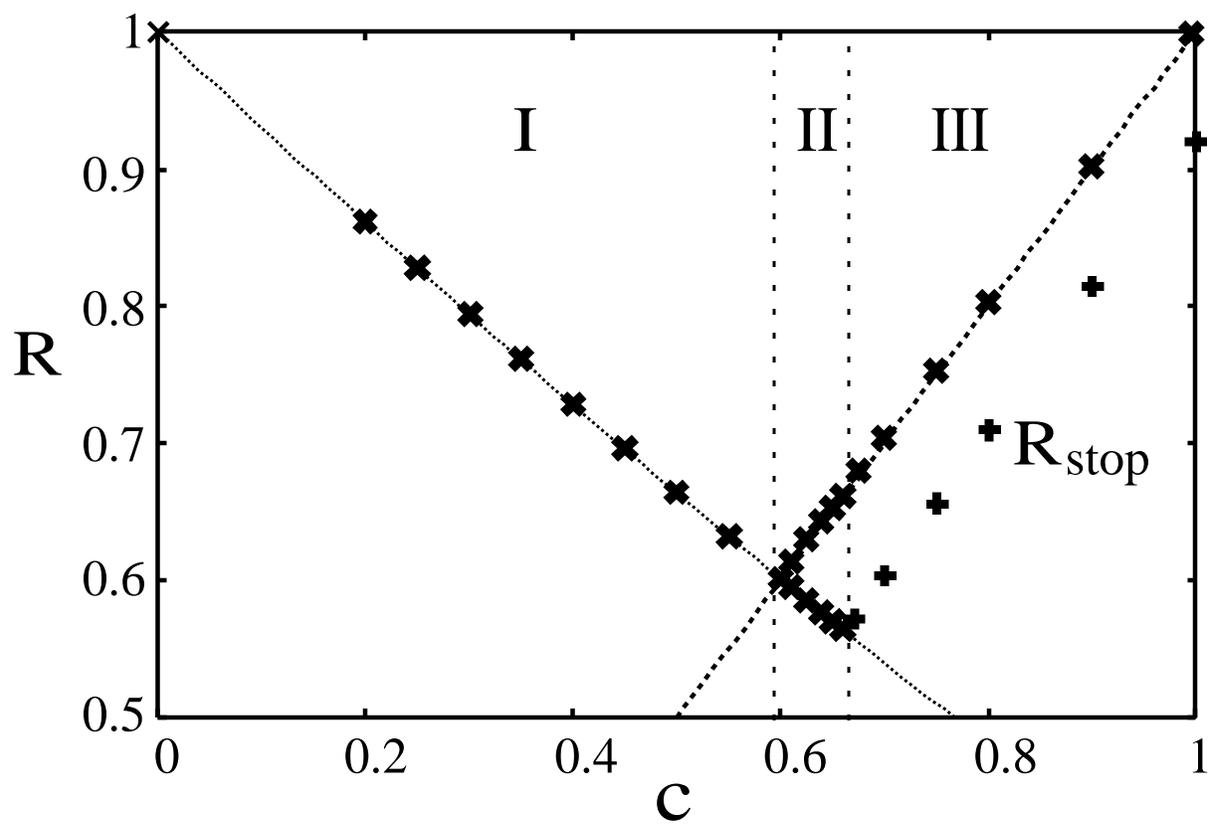
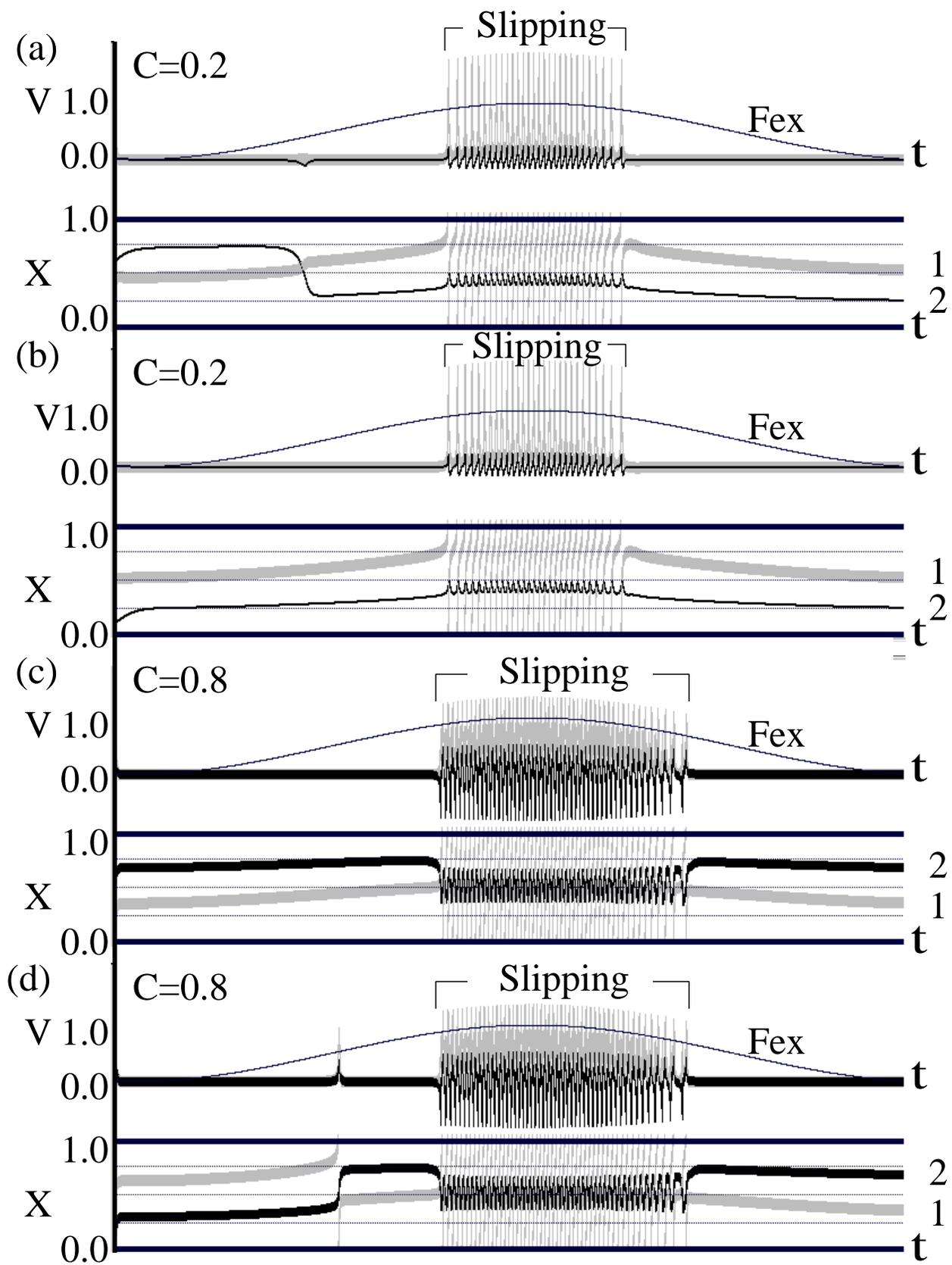
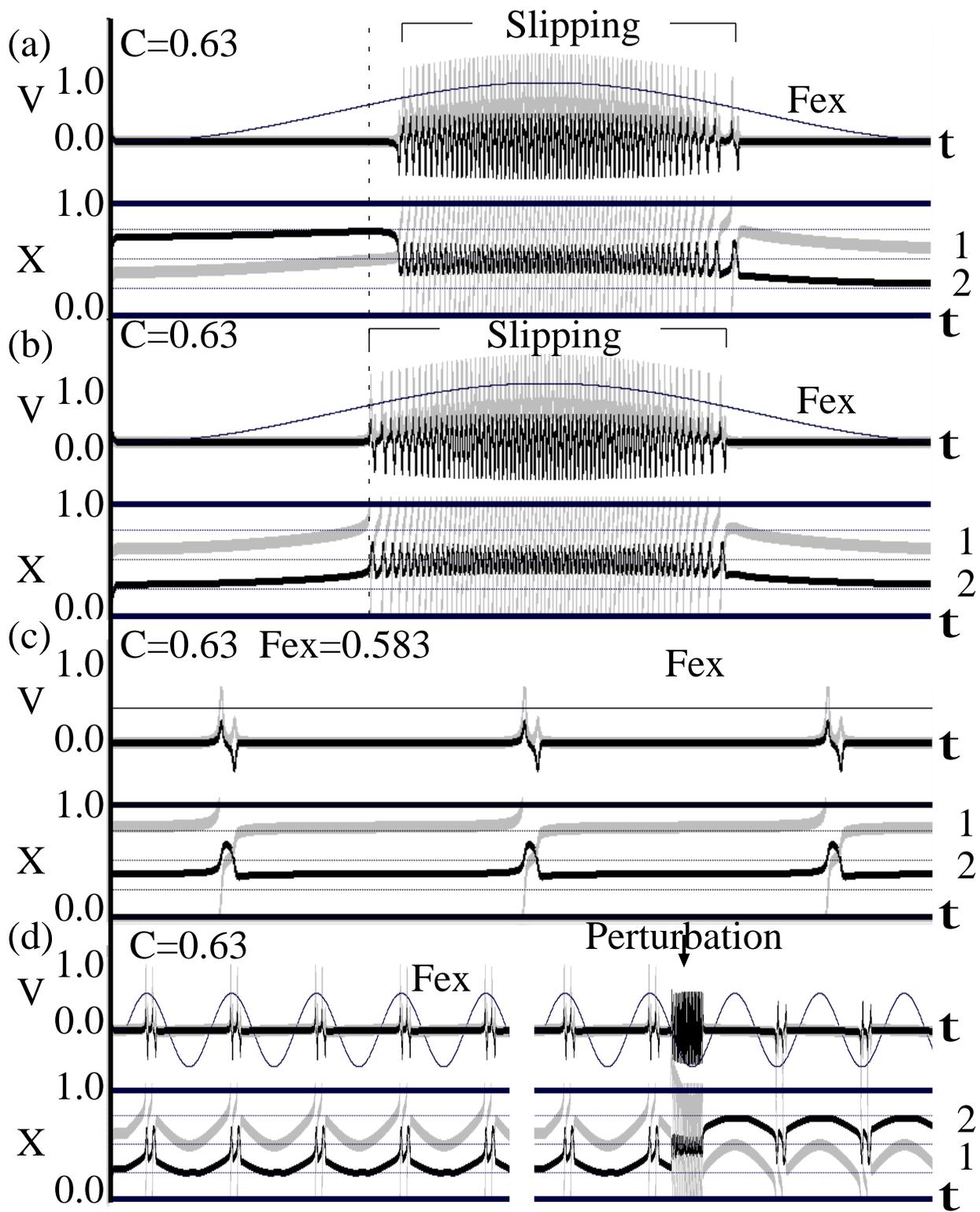
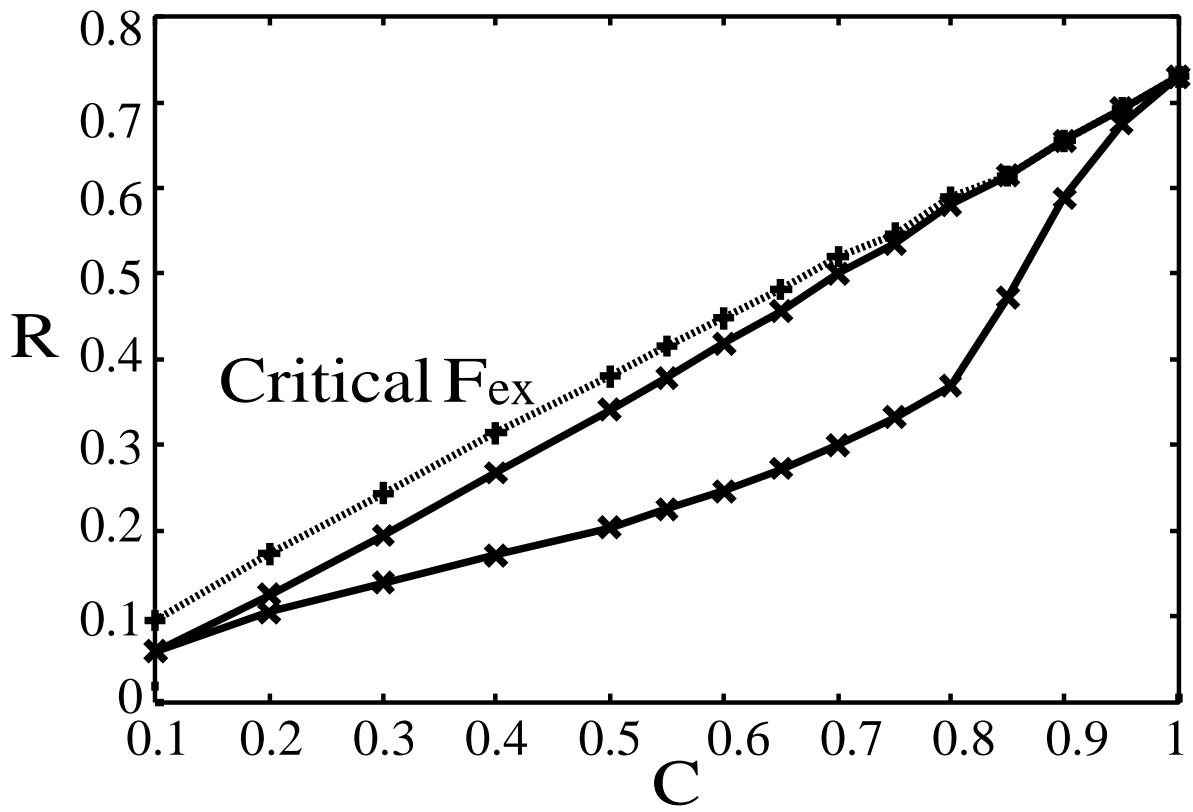
\end{document}